\documentclass[mathleft,fleqn,%
]{an}
\usepackage{graphicx}
\usepackage[varg]{txfonts}
\begin{document}

\Pagespan{1}{}
\Yearpublication{2014}%
\Yearsubmission{2014}%
\Month{0}%
\Volume{999}%
\Issue{0}%
\DOI{asna.201400000}%

\title{Gaia Benchmark stars and their twins in the Gaia-ESO Survey}

\author{Paula Jofr\'e\inst{1}\fnmsep\thanks{Corresponding author:
        {pjofre@ast.cam.ac.uk}}
}
\titlerunning{Gaia benchmark stars' twins}
\authorrunning{P. Jofr\'e}
\institute{
Institute of Astronomy, University of Cambridge, Madingley Road, CB3 0HA Cambridge, UK. 
}

\received{XXXX}
\accepted{XXXX}
\publonline{XXXX}


\abstract{%
The Gaia benchmark stars are stars with very precise stellar parameters that cover a wide range in the HR diagram at various metallicities. They are meant to be good representative of typical FGK stars in the Milky Way. Currently, they are used by several spectroscopic surveys to validate and calibrate the methods that analyse the data.  I review our recent activities done for these stars. Additionally, by applying our new method to find stellar twins  on the Gaia-ESO Survey, I  discuss how good representatives of Milky Way stars the benchmark stars are and how they distribute in space.}
\maketitle

\section{Introduction}
Many modern studies of stars aiming at understanding how our Galaxy formed and evolved consist in performing automatic and statistical analyses of large sample of stars. For instance, spectroscopic surveys are currently collecting high-resolution spectra of thousands of stars, such as RAVE (\cite{rave}), the Gaia-ESO Survey (\cite{ges}), the APOGEE Survey (\cite{apogee}) or the GALAH Survey (\cite{galah}). Numerous pipelines to extract the stellar information from these spectra, in particular, the star's effective temperature, surface gravity, metallicity and $\alpha$ abundances, are constantly under development.  

Most of these pipelines are designed to work on a specific instrument which has a given resolution and wavelength coverage.   For example,  for the RAVE survey the MATISSE pipeline is applied (\cite{2006MNRAS.370..141R}, \cite{2011AA...535A.106K}), which is designed to work in the Gaia RVS range (around the IR Ca II triplet $\lambda$8500 \AA).  For the analysis of the  Gaia-ESO Survey,  several pipelines designed to work in the optical range (between 4800 and 6800~\AA) and in the Gaia-RVS range are combined (see \cite{2014AA...570A.122S} and Recio-Blanco et al.~in prep, for details). For the APOGEE Survey, the FERRE pipeline (\cite{2015AAS...22542207A}) is applied to work on infrared spectra covering the range of $\sim15000$ to  17500~\AA\ of red giant stars. 

To evaluate how these pipelines work, and more importantly, to combine the stellar parameters of different datasets,  a sample of common stars is of extreme importance.  This sample needs to cover a wide range of temperatures, surface gravities and metallicities.  This is the purpose of the Gaia FGK benchmark stars: to provide  well-defined stellar parameters for set of typical survey FGK-type stars that can be used as calibrator pillars.

\begin{figure*}
\includegraphics[width=\linewidth]{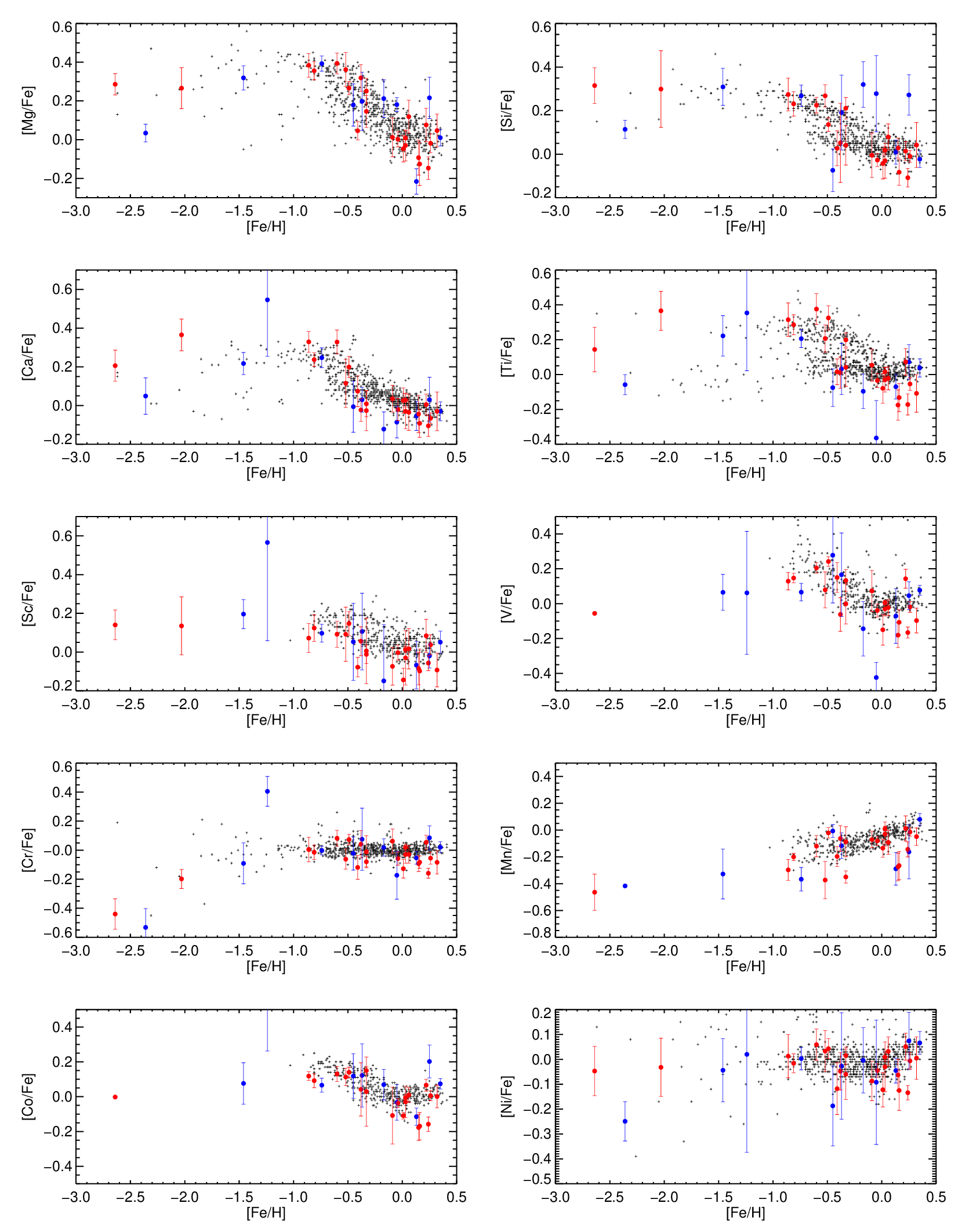}
\caption{Distribution of elements of 34 benchmark stars analysed in \cite{paperIV} as a function of metallicity. For illustration purposes,  the back crosses in the background show the distribution of the abundances from \cite{2014AA...562A..71B} and \cite{2015AA...577A...9B}. The blue symbols correspond to the ``not-recommended'' benchmarks, e.g., those declared as too uncertain in \cite{paperI} to be good reference stars, and those too cool ($\mathrm{T}_{\mathrm{eff}} < 4000$~K), which have abundances that need to be treated with care.}
\label{abus}
\end{figure*}

\section{Benchmark stars reference parameters}\label{req}

For a star to be ``granted'' as benchmark, its parameters need to be determined using information beyond its spectrum. Thus, it must satisfy the following requirements: 

\begin{enumerate}
\item {\it To be a typical Milky Way star}:  They need to have an ordinary spectrum from spectroscopic surveys. Only like this pipelines analysing these  spectra will be well calibrated. 
\item {\it To have an accurate parallax}:  Knowing the distance of reference stars is crucial for several reasons that will be seen throughout this article. 
\item {\it To have an angular diameter  measured from interferometry}:  The angular diameter is the key point. Combined with the parallax, the angular diameter gives a direct measurement of the radius of the star. 
\item {\it To have a measured bolometric flux}: This quantity, combined with the angular diameter, gives us a value for the effective temperature obtained from the Stefan-Boltzmann law. This temperature is independent of spectroscopy. 
\item {\it To have a well determined mass}: With radius in hand, surface gravity can be obtained from the Newton's law if we know the mass of the star. 
\item {\it To have a good spectrum}: A high-resolution and high signal-to-noise spectrum allows us to determine spectral parameters accurately. 
\end{enumerate}
In practice, these requirements imply that stars need to be close-by, very bright and with a large angular size. Smaller or fainter objects are not possible to resolve accurately with current interferometric instruments such as CHARA or AMBER.   The current sample of benchmark stars cover a magnitude range of $V \sim [0.1, 8]$~mag. This issue might present a challenge to many survey instruments, which are designed to target several faint objects simultaneously.  Gaia-ESO for example overcomes this problem by adopting a special observing procedure, with much shorter exposure times than the normal survey observations.

Because the aim is to have as many benchmark stars as possible, not all stars in our sample satisfy all the criteria enumerated above yet. Of the 34 initial sample stars, 7 do not have a measurement of angular diameter and 8 do not have a measurement of the bolometric flux so we use surface brightness relations instead. Ideally seismic targets or binaries are the best candidates to have a robust mass determination, but only 14 have seismic observations and 4 are in binary systems.  These measurements are used to calibrate the mass determination coming from evolutionary tracks. Furthermore, 2 stars with direct measurements of angular diameter or mass remain too uncertain to retrieve consistent atmospheric parameters.   For details on this see \cite{paperI}, who extensively discusses the determination of effective temperature and surface gravities of the initial sample of benchmark stars.  In summary,  having a large sample of benchmark stars satisfying the criteria mentioned above covering well the parameter space of typical FGK stars is quite challenging. 

\subsection{Spectroscopic parameters}

The best way to derive the metallicity of stars is from the spectrum.  Since the benchmark stars are very bright objects, they have been extensively studied in the past by several different authors. However, as discussed in \cite{paperIII}, the metallicity values of benchmark stars found in the literature are highly inhomogeneous. This is because the different authors use different methods, input data and scales for the effective temperature, surface gravity or the solar abundances. Differences of up to 1~dex can be found for the metallicity in some stars. Furthermore, there is no work in the literature that has analysed all the benchmark stars homogeneously. This is not surprising, the benchmark stars cover more than 3000~K in temperature, 2~dex in metallicity and have gravities from 0 to 5. Works in the literature are normally focused on one type of star only, i.e., giants or dwarfs.  It is thus important to re-determine metallicity (and elemental abundances) in an homogeneous fashion, which was presented in \cite{paperIII}. For that, we collected high resolution and high signal-to-noise archive spectra from UVES, HARPS, NARVAL and ESPaDOnS,  and compiled a dedicated library of high-resolution and high signal-to-noise spectra such that every spectrum would have the same format and resolution.  This library is described in \cite{paperII}. 

Recently, motivated by the fact that automatic pipelines analysing spectroscopic surveys not only determine stellar parameters but also elemental abundances, we determined  abundances for the benchmark stars of ten different elements. In a similar way than the Fe abundance, we provide reference values for the $\alpha$ elements Mg, Si, Ca and Ti, and the iron-peak elements Sc, V, Cr, Mn, Co and Ni.   An extensive discussion on this can be found in \cite{paperIV}.    Figure.~\ref{abus}  shows with filled circles the distribution of elements as a function of metallicity as taken from Jofr\'e et al.~(2014, 2015a) for the benchmark stars.  The error bars correspond to the line-to-line uncertainty. For illustration purposes,  the crosses in the background correspond to the abundances of the samples of \cite{2014AA...562A..71B} and \cite{2015AA...577A...9B}.  

 The figure tells us that the benchmark stars are well distributed in $\alpha-$metallicity space, with low-$\alpha$ stars attributed to thin disk stars, high-$\alpha$ stars to thick disk metal-rich stars, and few metal-poor stars attributed to halo or the metal-poor thick disk stars.  The good benchmark stars (in red) have accurate abundances following the typical Galactic trends and dispersions.

\subsection{Products and current activities}\label{current}
The spectral library and the parameters of the Gaia benchmark stars are currently being used for calibration purposes within Gaia-ESO, GALAH, APOGEE and also within some activities of the DPAC-CU8 for Gaia. Since they are becoming an excellent database to link the different spectroscopic surveys, we are dedicating great efforts to improve the sample in all possible aspects.  

\begin{figure}
\includegraphics[width=\linewidth]{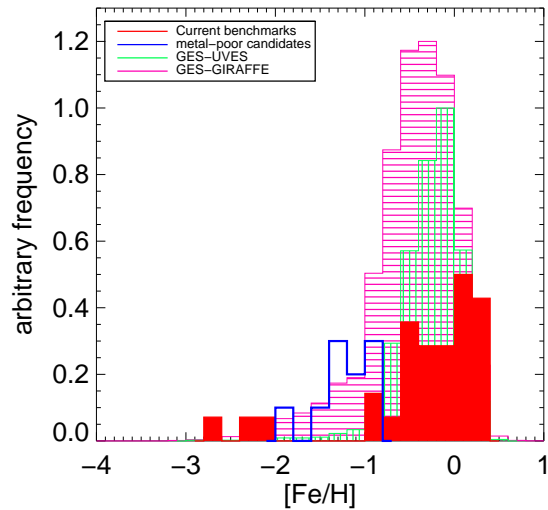}
\caption{Metallicity distribution Gaia-ESO DR2 for UVES and GIRAFFE and the benchmark stars. }
\label{met}
\end{figure}

One of the deficiencies can be noted from Fig.~\ref{abus}, and is well represented in Fig.~\ref{met}: There are no good benchmark stars with $-2 < \mathrm{[Fe/H]} < -1$.  Figure~\ref{met} shows in red the metallicity distribution of the 34 current benchmark stars, where the gap can be clearly seen. On the background, the metallicity distribution from Gaia-ESO DR2 parameters is shown with green vertical lines for the UVES sample and with pink horizontal lines for the GIRAFFE sample. It is possible to see that a large fraction of thick disk stars observed with GIRAFFE are not well represented with the Gaia benchmark stars. The blue open histogram represents the sample of new metal-poor candidates (Hawkins et al, in prep).  

Additionally, as mentioned before, there are angular measurements that are uncertain or not available.  There are several observational campaigns (P.I. Karovicova/Creevey) to improve the benchmark stars sample with better angular diameters.

\section{Twins of benchmark stars in Gaia-ESO}
We may ask the question of how good representatives the benchmark stars really are (how well is Point~1 of Sect.~\ref{req} satisfied). That is, how many stars can we find in a spectroscopic survey that have equal spectra to the benchmark stars? 
By using the method presented in \cite{twins}, the spectra of benchmark stars as taken with the UVES instrument in the Gaia-ESO Survey were compared against the UVES iDR4 dataset of the Milky Way field.  Briefly, we analysed the spectral lines used in the spectral analyses of the benchmark stars in \cite{paperIII} and in \cite{paperIV}, as these lines are good clean lines for typical FGK stars. Twins were found by comparing the equivalent widths (EWs) of lines between two stars as a function of the excitation potential (EP) of the line. Those  cases showing no trend or offset of EW vs EP were assigned to be twins.

Only twins of benchmark stars dwarfs were found, which is not surprising since the UVES part of the survey targets mostly dwarfs. More specifically, the stars 18~Sco and $\mu$~Ara were found to have the largest number of twins, while the hot dwarf Procyon had none.  

\begin{figure}
\includegraphics[width=\linewidth]{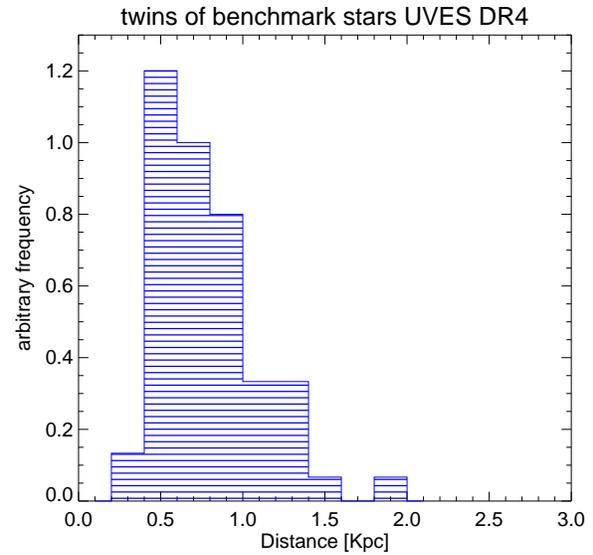}
\caption{Distance distribution of benchmark stars' twins in Gaia-ESO.}
\label{ges_twins}
\end{figure}

Furthermore, as shown in \cite{twins}, spectroscopic twin stars can be used to determine very precise and model-independent distances, if they have good photometry and one of them has an accurate parallax.  Using the very precise parallaxes of the benchmark stars, the distance distribution of their twins found in the Gaia-ESO UVES sample is illustrated in Fig.~\ref{ges_twins}. The distances were calculated considering 2MASS photometry.  The stars are located within 1.5 Kpc, agreeing with the model-dependent distance measurements of \cite{maria} for the UVES DR1 sample of Gaia-ESO.  

\section{Summary}
 Now the Gaia satellite is orbiting in space, collecting data for the largest and most accurate 3D stellar map of our history. The spectra of millions of stars yet unknown as observed by Gaia, Gaia-ESO, GALAH, APOGEE, or any other future survey, will be analysed and parametrised according to calibration samples. With our dedicated documentary work on their atmospheric properties and spectral line information, the Gaia benchmark stars provide fundamental material to connect these surveys and to contribute to a better understanding of our home galaxy.

\acknowledgements
  This work was partly supported by the European Union FP7 programme through ERC grant number 320360. It is a pleasure to thank numerous colleagues that collaborate actively on the projects presented here. In particular, the author acknowledges U. Heiter, C. Soubiran, S. Blanco-Cuaresma, G. Gilmore, T. M\"adler, A. Casey, T. Masseron, K. Hawkins, and the Gaia-ESO collaboration, for substantial contribution in work related to this article.

\end{document}